\documentclass[tightenlines,superscriptaddress,twocolumn,prd,notitlepage,nofootinbib,preprintnumbers]{revtex4-1}

\usepackage{graphicx}
\usepackage{amsmath}
\usepackage{amsfonts}
\usepackage{amssymb}
\usepackage{float}
\usepackage{hyperref}
\usepackage{multirow}
\usepackage{cancel}
\usepackage{xcolor}
\usepackage{soul}
\usepackage{makecell}
\usepackage{minibox}
\usepackage[utf8]{inputenc}

\begin{document}

% page numbers bottom-center
\pagestyle{plain}

%%%%%%%%%%%%%%%%%%%%%%%%%%%%%%%%%%%%%%%%%%%%%%%%%%%%%%%%%%%%%%%%%%%%%%%%%%%%
\preprint{LA-UR-16-26969}
\title{Developing diagnostic tools for low-burnup reactor samples}

\author{Patrick Jaffke}
\email[corresponding author: ]{pjaffke@lanl.gov}
\affiliation{Los Alamos National Laboratory, Los Alamos, NM 87545, USA}
\author{Benjamin Byerly}
\affiliation{Los Alamos National Laboratory, Los Alamos, NM 87545, USA}
\affiliation{Department of Geology and Geophysics, Louisiana State University, Baton Rouge, LA 70803, USA}
\author{Jamie Doyle}
\affiliation{Los Alamos National Laboratory, Los Alamos, NM 87545, USA}
\author{Anna Hayes}
\affiliation{Los Alamos National Laboratory, Los Alamos, NM 87545, USA}
\author{Gerard Jungman}
\affiliation{Los Alamos National Laboratory, Los Alamos, NM 87545, USA}
\author{Steven Myers}
\affiliation{Los Alamos National Laboratory, Los Alamos, NM 87545, USA}
\author{Angela Olson}
\affiliation{Los Alamos National Laboratory, Los Alamos, NM 87545, USA}
\author{Donivan Porterfield}
\affiliation{Los Alamos National Laboratory, Los Alamos, NM 87545, USA}
\author{Lav Tandon}
\affiliation{Los Alamos National Laboratory, Los Alamos, NM 87545, USA}

\date{\today}

%%%%%%%%%%%%%%%%%%%%%%%%%%%%%%%%%%%%%%%%%%%%%%%%%%%%%%%%%%%%%%%%%%%%%%%%%%%%
\begin{abstract}
We test common fluence diagnostics in the regime of very low burnup natural uranium reactor samples. The fluence diagnostics considered are the uranium isotopics ratios $^{235}$U/$^{238}$U and $^{236}$U/$^{235}$U, for which we find simple analytic formulas agree well with full reactor simulation predictions. Both ratios agree reasonably well with one another for fluences in the mid $10^{19}\,\mathrm{n/cm^2}$ range. However, below about $10^{19}\,\mathrm{n/cm^2}$ the concentrations of $^{236}$U are found to be sufficiently low that the measured $^{236}$U/$^{235}$U ratios become unreliable. We also derive and test diagnostics for determining sample cooling times in situations where very low burnup and very long cooling times render many standard diagnostics, such as the $^{241}$Am/$^{241}$Pu ratio, impractical. We find that using several fragment ratios are necessary to detect the presence of systematic errors, such as fractionation.
\end{abstract}

\maketitle

%%%%%%%%%%%%%%%%%%%%%%%%%%%%%%%%%%%%%%%%%%%%%%%%%%%%%%%%%%%%%%%%%%%%%%%%%%%
\section{Isotopics Introduction\label{sec:Intro}}

Determining the reactor environment that a particular spent fuel sample
experienced is critical information for non-proliferation and reactor
verification. In particular, the fluence is often related to the fuel
burnup and, hence, the plutonium production and grade~\cite{Stepanov1980}.
This makes the fluence an important parameter for nonproliferation and arms
reduction~\cite{Wood:2002}. The fluence of a sample can be inferred in many ways,
but is most commonly derived from isotopic ratios of actinides,
such as $^{235}$U/$^{238}$U or $^{236}$U/$^{235}$U~\cite{Boulyga:2002,Boulyga:2006}
and various plutonium ratios~\cite{Kim2015924}. Additional methods utilize the
ratios of activated isotopes in cladding and moderator material, such as the graphite isotope ratio method (GIRM)~\cite{Reid:1999,Gesh:2004,Gasner:2011}, or of ratios of long-lived fragments such as cesium~\cite{Caruso:2007,Ansari:2007,Kim2015924}, europium~\cite{Caruso:2007}, or neodymium~\cite{Kim2015924}. The cooling time is often determined with ratios utilizing short-lived actinides, such as $^{241}$Pu/$^{241}$Am~\cite{Mayer:2012}, but can also be inferred by gamma spectroscopy of fragments~\cite{Gauld:2012}. The cooling time provides one with an estimate of the sample age, which is also pertinent for forensics and nonproliferation.

One can determine the final activities, abundances, and ratios of nuclides with detailed reactor simulations, provided a burnup history and initial fuel composition. Our goal is to invert this process, where one begins with measured isotopic abundances or ratios and then determines the reactor parameters, such as the fluence and cooling time. We focus on these two parameters as they are derived from so-called linear systems, which have simpler analytical forms, in the low burnup regime. Non-linear systems can be used to infer parameters, such as the flux and shutdown history~\cite{Hayes:2012sg}. In our regime of interest, new cooling time diagnostics are developed and verified alongside the standard fluence diagnostics. Several cooling time diagnostics are utilized to detect the presence of systematic errors. We used low burnup archived samples, available at Los Alamos National Laboratory, to test these diagnostics. The chemical analyses to determine the abundances of the actinides and fission fragments for our low burnup samples can be found in Ref.~\cite{Byerly:2015} and Ref.~\cite{Tandon:2009}.

This paper is structured as follows. The fluence diagnostics are discussed in Sec.~\ref{sec:Fluence}. Cooling time diagnostics are discussed and derived in Sec.~\ref{sec:CoolingTime}. The diagnostics are verified with reactor simulations and theoretical errors are generated in Sec.~\ref{sec:Verification}. The diagnostics are then applied to low burnup reactor samples to determine their fluence, cooling time, and sensitivity to systematic errors in Sec.~\ref{sec:Application}. We conclude in Sec.~\ref{sec:Conclusion}.

%%%%%%%%%%%%%%%%%%%%%%%%%%%%%%%%%%%%%%%%%%%%%%%%%%%%%%%%%%%%%%%%%%%%%%%%%%%
\section{Fluence Diagnostics\label{sec:Fluence}}
The fluence diagnostics considered in this work utilize the uranium
isotopic ratios: $^{235}$U/$^{238}$U and $^{236}$U/$^{235}$U. Ratios utilizing
moderator materials require a sample removal from the existing reactor, which
is often not feasible or impacts future reactor design and safety. In addition,
some long-lived fragments, such as $^{134}$Cs or $^{154}$Eu, are produced in
extremely low concentrations for low burnup scenarios creating experimental
difficulties. Finally, $^{239}$Pu cannot be used, as its accumulation is not
precisely linear in fluence in low burnup scenarios, thus displaying a
flux dependence. For these reasons, we focus on the uranium ratios above which
are trivially related to the fluence via
%%%%%%%%%%%%%%%%%%%%%%%%%%%%%%%%
\begin{equation}
\begin{split}
\epsilon(\Phi,\epsilon_0) &= \epsilon_0 e^{-\Phi(\sigma^T_{U235} - \sigma^T_{U238})} \\
\rho(\Phi) &= \bigg( \frac{\sigma^c_{U235}}{\sigma^T_{U236} -\sigma^T_{U235}} \bigg) \bigg( 1-e^{-\Phi (\sigma^T_{U236} - \sigma^T_{U235})} \bigg).
\end{split}
\label{eq:Uratios}
\end{equation}
%%%%%%%%%%%%%%%%%%%%%%%%%%%%%%%%
Here, $\epsilon$ denotes the $^{235}$U/$^{238}$U ratio and $\rho$ the
$^{236}$U/$^{235}$U ratio. The superscripts on the cross-sections $\sigma$ are for
capture ($c$) or total ($T$) reactions and $\Phi$ is the fluence
~\footnote{In addition, one would group $\Phi$ and $\sigma$ by energy-groups, but
we list the $1$-group values for simplicity.}.

We immediately note that $\epsilon$ depends on the initial
ratio $\epsilon_0$. This implies that a measurement of $\Phi$ via the
$^{235}$U/$^{238}$U ratio is only valid when the initial enrichment is known.
In the case of our low burnup samples, all indicated natural uranium (NU) as
the initial fuel~\cite{Byerly:2015}. On the other hand, the determination of
$\Phi$ from $\rho$ is insensitive to the initial fuel, but requires
a measurement of $^{236}$U, which is produced in very low quantities
when the burnup is low. A final note is that a measurement of
$\Phi$ using Eq.~\ref{eq:Uratios} will be most sensitive to the thermal fluence,
as these cross-sections dominate (specifically $^{235}$U thermal fission).

Inverting Eq.~\ref{eq:Uratios} produces the fluence diagnostics we will apply to
the low burnup samples
%%%%%%%%%%%%%%%%%%%%%%%%%%%%%%%%
\begin{equation}
\begin{split}
\Phi &= \frac{\ln (\epsilon_0/\epsilon)}{\sigma_{U235}^T - \sigma_{U238}^T} \\
\Phi &= \frac{1}{\sigma_{U235}^T - \sigma_{U236}^T} \ln \bigg( \frac{\sigma_{U235}^c - \rho (\sigma_{U236}^T - \sigma_{U235}^T)}{\sigma_{U235}^c} \bigg).
\label{eq:FluenceD}
\end{split}
\end{equation}
%%%%%%%%%%%%%%%%%%%%%%%%%%%%%%%%
Measurement of the values of $\epsilon$ or $\rho$ are typically accomplished by
chemical separation~\cite{Natsume:1972,Abernathey:1972,Marsh:1974}, followed by
gamma spectroscopy~\cite{Kim2015924}, thermal ionization mass spectrometry
(TIMS)~\cite{Byerly:2015}, or inductively coupled plasma mass spectrometry
(ICP-MS)~\cite{Boulyga:2002,Boulyga:2006}. Specifically, the measurement of
$^{236}$U is made difficult as isobaric interferences arise when small quantities
of $^{236}$U exist amidst large $^{235}$U and $^{238}$U quantities. Additionally,
the $\alpha$-decay peak of $^{235}$U can interfere in a $^{236}$U/$^{238}$U
measurement done via $\alpha$-spectrometry~\cite{SANCHEZ1992219,Iturbe:1992}.

%%%%%%%%%%%%%%%%%%%%%%%%%%%%%%%%%%%%%%%%%%%%%%%%%%%%%%%%%%%%%%%%%%%%%%%%%%%
\section{Cooling Time Diagnostics\label{sec:CoolingTime}}
Cooling time diagnostics must be selected specifically for the context of low
burnup samples. For example, the $^{241}$Pu/$^{241}$Am ratio cannot be used as
neither $^{241}$Pu nor $^{241}$Am are produced in appreciable amounts at low burnups.
Similar issues preclude the use of other unstable actinides from NU fuel or
$^{134}$Cs and $^{154}$Eu, both of which are non-linear nuclides~\cite{Huber:2015ouo}
that rely on neutron capture as their primary production channel. This indicates
that common cooling time diagnostics that utilize same-species ratios to avoid
fractionation~\cite{Freiling:1961}, such as the $^{134}$Cs/$^{137}$Cs
ratio~\cite{Navarro:2011}, are invalid. In addition, the special case of extremely
long cooling times~\footnote{We define the cooling time as the sum of all pure
decay periods, including shutdowns.} ($\sim 20\,\mathrm{yr}$) invalidate the use of
some major decay heat tags, such as $^{106}$Ru and $^{144}$Ce~\cite{Bergelson:2005}.
Thus, the cooling time diagnostic requires nuclides that are appreciably produced in
low burnup scenarios, have long half-lives, and are easy to chemically separate and
analyze. These requirements naturally lead one to the so-called `linear' fission
fragments, described by:
%%%%%%%%%%%%%%%%%
\begin{enumerate}
\item The linear fragment $N_L$ has a halflife such that its decay constant $\lambda_L$ satisfies $\lambda_L T_\mathrm{irr} \ll 1$.
\item The fission product cumulative yields for $N_L$ are large.
\item The beta-parents of $N_L$ have halflives such that they are in equilibrium during $T_\mathrm{irr}$.
\end{enumerate}
%%%%%%%%%%%%%%%%%

These fragments are dubbed `linear' as their production is linear in the fluence
$\Phi=\phi T_\mathrm{irr}$. The first criteria ensures that the fragment is long-
lived relative to the irradiation period of the reactor. The second criteria demands
that the fragment is appreciably produced in fission. The third criteria allows one
to derive a simple analytical expression for $N_L$, independent of its $\beta$-
parents. For our low burnup purposes, $^{85}$Kr, $^{125}$Sb, $^{137}$Cs, and
$^{155}$Eu are linear fragments. Next, we proceed to derive the simple expressions
for these and verify that they satisfy the criteria above with detailed reactor
simulations.

All nuclides in a reactor environment are governed by depletion equations, which
form the basis for constructing an interaction matrix between the various nuclides.
This is the structure utilized by many reactor simulation
codes~\cite{SCALEmanual,MURE:2009}, which often solve these massive
($\sim 2000$ species) systems as an eigenvalue problem~\cite{Pusa:2010}. In
our case, we can utilize linear fragments to construct a simple isolated system,
which resembles a Bateman equation~\cite{Bateman:1910},
%%%%%%%%%%%%%%%%%%%%%%%%%%%%%%%%
\begin{equation}
\frac{dN_L}{dt} = - \tilde{\lambda}_{L}N_{L} + \vec{Z}_{L} \cdot \vec{\mathcal{F}}.
\label{eq:SimpleBateman}
\end{equation}
%%%%%%%%%%%%%%%%%%%%%%%%%%%%%%%%
The positive (negative) terms denote production (depletion) channels and we use an
effective decay constant $\tilde{\lambda}=\lambda+\phi\sigma^T$. We note that the
full depletion equation, which resembles Eq.~1 of Ref.~\cite{Isotalo2011261},
reduces to Eq.~\ref{eq:SimpleBateman} after applying $\lambda_i\gg b_{j,i}\lambda_j$
(criteria $3$), noting that $\sigma_{j,i}\phi\ll b_{j,i}\lambda_j$ is satisfied for
most fragments~\footnote{An exception to this is the $^{135}$Xe$(n,\gamma)^{136}$Xe
cross-section.}, and adding an explicit fission term. Thus, Eq.~\ref{eq:SimpleBateman}
states that a linear fragment $N_L$ is produced in fission with a fission rate
vector $\vec{\mathcal{F}}$ and a cumulative yield $\vec{Z}_L$, both of which span
the major fissiles, and is depleted through its decay and neutron-capture.

\begin{figure}[t]
\centering
\includegraphics[width=\columnwidth]{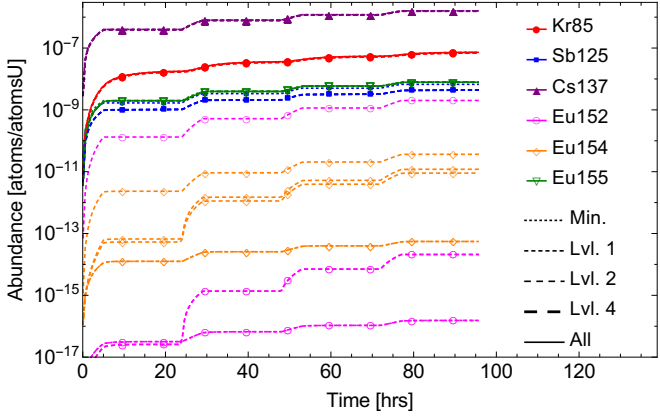}
\caption{\label{fig:AddingData} Relative abundances of several fission
  fragments from a simulation of an irradiation history consisting of
  $4$ $\lbrack 5,19\rbrack\mathrm{hr}$ ON/OFF periods and a thermal
  flux of $\phi_t=8.5\times10^{13}\,\mathrm{n/cm^2/sec}$ beginning
  with natural uranium. The simulation uses CINDER08~\cite{CINDER08}
  cross-sections and the decay data was allowed to vary between
  ENDF7~\cite{Chadwick:2011xwu}, JEFF~\cite{JEFF311}, and
  JENDL~\citep{JENDL}, with no observed difference. Linear fragments
  show no dependence on the layer of nuclear data. Color online.}
\end{figure}

Solving Eq.~\ref{eq:SimpleBateman} yields
%%%%%%%%%%%%%%%%%%%%%%%%%%%%%%%%
\begin{equation}
N_L(t) = \bigg( {N_L}_0 - \frac{\vec{Z}_L \cdot \vec{\mathcal{F}}}{\tilde{\lambda}_L}
  \bigg) e^{-\tilde{\lambda}_L t} +  \frac{\vec{Z}_L \cdot \vec{\mathcal{F}}}{\tilde{\lambda}_L},
\label{eq:LinearBateman}
\end{equation}
%%%%%%%%%%%%%%%%%%%%%%%%%%%%%%%%
with an initial nuclide abundance ${N_L}_0$. We note that most linear fragments
satisfy $\tilde{\lambda}_L \approx\lambda_L$, which can be verified by solving for
the critical flux when decay and neutron channels have equal rates. Standard
reactor fluxes are far below the critical fluxes of most fragments, ensuring that
decay channels dominate. An exception to this is $^{155}$Eu and, in high thermal
flux reactors~\cite{Ilas:2011}, $^{85}$Kr. We include the effective decay constant
in our derivations for generality. One can easily verify that our selected
fragments are linear in nature using reactor simulations. We use a finite-difference
methods solver for the interaction matrix, where the included nuclear
data can be varied. A sample irradiation history is given by $4$ cycles of
$\lbrack 5,19\rbrack\mathrm{hr}$ ON/OFF periods and a thermal flux of
$\phi_t=8.5\times10^{13}\,\mathrm{n/cm^2/sec}$. The resulting relative abundances
for our linear fragments and, for comparison, two non-linear fragments
($^{152,154}$Eu) are shown in Fig.\ref{fig:AddingData}.
%%%%%%%%%%%%%%%%%%%%%%
%%%%%%%%%%%%%%%%%%%%%%

The minimum layer of nuclear data considered just our fragment of interest (FOI).
This physically represents the case when each FOI is given by
Eq.~\ref{eq:LinearBateman}. Layer $1$ added the $\beta$-parents. Layer $2$ added
the $(n,\gamma)$ parent. Layer $4$ included the primary, secondary, and (in some
cases) tertiary $(n,\gamma)$ channels as well as all of their $\beta$-parents
with halflives greater than $30\,\mathrm{sec}$. We also included a simulation of
all nuclides with available data ($\sim700$). From Fig.~\ref{fig:AddingData}, one
can verify that $^{85}$Kr, $^{125}$Sb, $^{137}$Cs, and $^{155}$Eu are linear as
they have very little dependence on the layer of nuclear data and, thus, are
accurately given by Eq.~\ref{eq:LinearBateman}. None of the fragments studied
varied significantly between the major fission yields
libraries~\cite{Chadwick:2011xwu,JEFF311,JENDL}.

To derive the cooling time diagnostic, we first expand Eq.~\ref{eq:LinearBateman}
with $\tilde{\lambda}T_\mathrm{irr}\ll 1$ (criteria $1$) and arrive at
%%%%%%%%%%%%%%%%%%%%%%%%%%%%%%%%
\begin{equation}
N_L(t,T_c)= \Phi(\vec{Z}_L\cdot \vec{\Sigma}_{\mathrm{fiss}})e^{-\lambda_L T_c}
  + \mathcal{O}((\tilde{\lambda}_L T_{\mathrm{irr}})^2),
\label{eq:SimpleLinearSol}
\end{equation}
%%%%%%%%%%%%%%%%%%%%%%%%%%%%%%%%
once we have set ${N_L}_0 = 0$, accounted for the decay after a cooling time $T_c$,
and separated the fission rate vector into a weighted fission cross-section and
the flux through the relation $\vec{\mathcal{F}}=\vec{\Sigma}_\mathrm{fiss}\phi$.
The expansion to arrive at Eq.~\ref{eq:SimpleLinearSol} is easily valid for all
fragments used here, except $^{155}$Eu which deviates from it by $\sim1-3\%$ due to
its large cross-section. As $\vec{\Sigma}_\mathrm{fiss}$ is the fission
cross-sections weighted by the fissile abundances, one can determine
$\vec{\Sigma}_\mathrm{fiss}$ with similar chemical analyses as those used for the
fragments~\cite{Byerly:2015}.

\begin{figure}[h]
\setlength{\unitlength}{0.13in}
\centering
\begin{picture}(28,6)
\put(0,2.5){\color{black}\line(0,1){1.5}}
\put(2.5,2.5){\color{black}\line(0,1){1}}
\put(4.5,2.5){\color{black}\line(0,1){1.5}}
\put(6.2,2.5){\color{black}\line(0,1){1}}
\put(8,2.5){\color{black}\line(0,1){1.5}}
\put(10,3){$\ldots$}
\put(12,3){\color{black}\vector(1,0){14}}
\put(0.75,3.5){$\beta_1t$}
\put(2.85,3.5){$\gamma_1\tau$}
\put(4.75,3.5){$\beta_2t$}
\put(6.45,3.5){$\gamma_2\tau$}
\put(13.5,4.5){$(1-\displaystyle\sum_k^{N-1}\beta_k)t$}
\put(20,4.5){$(1-\displaystyle\sum_k^{N-1}\gamma_k)\tau$}
\put(0.7,1){Cycle $1$}
\put(4.75,1){Cycle $2$}
\put(17,1){Cycle $N$}
\put(0,3){\color{black}\line(1,0){9}}
\put(12.5,2.5){\color{black}\line(0,1){1.5}}
\put(20,2.5){\color{black}\line(0,1){1}}
\end{picture}
\caption{\label{fig:TimeDist} Generalized irradiation history with $N$
  cycles, each consisting of an irradiation time of length $\beta_i t$
  and cooling time $\gamma_i \tau$ with multiplicative factors $\beta_i$
  and $\gamma_i$ that sum to unity. The $N^{\mathrm{th}}$ cycle is the
  remainder of the total irradiation time $t$ and cooling time $\tau$
  with time ascending from left to right.}
\end{figure}
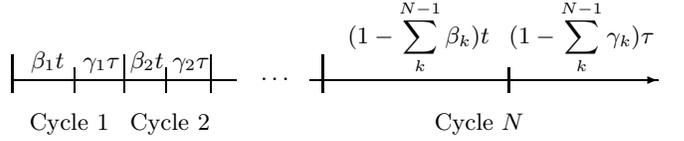

Universally setting ${N_L}_0=0$ appears to exclude cases with multiple irradiation
cycles. Suppose we have a distribution of irradiation and cooling times described
in Fig.~\ref{fig:TimeDist}, where $t$ and $\tau$ are the total irradiation and
cooling times across all cycles.
%%%%%%%%%%%%%%%%%%%%%%
%%%%%%%%%%%%%%%%%%%%%%
We recursively insert Eq.~\ref{eq:LinearBateman} into itself as an initial condition
for the following irradiation and cooling period to verify that distributing the
total irradiation and cooling time in a generalized way is a negligible effect on
our linear fragments. We find that the final activity ($\alpha_L=\lambda_LN_L$) of a
purely linear fragment (i.e. $\tilde{\lambda}_L = \lambda_L$) with a generic
distribution of $t$ and $\tau$ over $N$ cycles is given by
%%%%%%%%%%%%%%%%%%%%%%%%%%%%%%%%
\begin{equation}
\begin{split}
\alpha_L(t,\tau,\vec{\beta},\vec{\gamma},{N_L}_0) =& (\lambda_L {N_L}_0 -
  \vec{Z}_L \cdot \vec{\mathcal{F}}) e^{-\lambda_L (t+\tau)} \\&+ (\vec{Z}_L
  \cdot \vec{\mathcal{F}}) e^{-\lambda_L \tau} \times f(\vec{\beta},\vec{\gamma}),
\label{eq:NCycle}
\end{split}
\end{equation}
%%%%%%%%%%%%%%%%%%%%%%%%%%%%%%%%
with a pre-irradiation initial abundance ${N_L}_0$ and the function
$f(\vec{\beta},\vec{\gamma})$ is given as a sum and product of exponentials over
the additional $N-1$ cycles
%%%%%%%%%%%%%%%%%%%%%%%%%%%%%%%%
\begin{equation}
\begin{split}
&f(\vec{\beta},\vec{\gamma}) = \bigg( \displaystyle\prod_{i=1}^{N-1} e^{\lambda_L \gamma_i \tau} \bigg)
  + e^{-\lambda_L t} \times \\& \displaystyle\sum_{i=1}^{N-1} \bigg[ \bigg( \displaystyle\prod_{j=1}^i
  e^{\lambda_L \beta_j t} \bigg) \bigg( \displaystyle\prod_{k=1}^{i-1} e^{\lambda_L \gamma_k \tau} \bigg)
  - \bigg( \displaystyle\prod_{j=1}^i e^{\lambda_L (\beta_j t + \gamma_j \tau)} \bigg) \bigg].
\label{eq:FFactor}
\end{split}
\end{equation}
%%%%%%%%%%%%%%%%%%%%%%%%%%%%%%%%
This complex function for $N$ cycles reduces to unity when $N=1$. One can show that
criteria $1$, and the fact that the individual elements of $\vec{\beta}$ and
$\vec{\gamma}$ are limited by unitarity, restricts Eq.~\ref{eq:FFactor} to very
small deviations from $1$. We analyzed generic values for $\vec{\beta}$ and
$\vec{\gamma}$ within our expected $t$ and $\tau$ ranges and found that
Eq.~\ref{eq:FFactor} is well-constrained to $\lesssim1\%$ deviations from unity. An
exception to this is $^{125}$Sb, which showed larger deviations when the decay time
is concentrated towards earlier cycles (i.e. when $\gamma_1\gg\gamma_{k>1}$), but
this is disfavored for our samples. As no fragments are expected in pre-irradiated
fuel, we determine that ${N_L}_0=0$ is a valid assumption at the start of
irradiation and any subsequent cooling time diagnostic will now include intermediate
shutdowns.

With ${N_L}_0=0$, the final abundance of a linear fragment can be expressed as in
Eq.~\ref{eq:SimpleLinearSol}. A ratio of the activities of two linear fragments
removes the explicit dependence on $\Phi$ and creates the expression
%%%%%%%%%%%%%%%%%%%%%%%%%%%%%%%%
\begin{equation}
\alpha_{n,d}(\vec{\Sigma}_\mathrm{fiss},T_c) = \frac{\lambda_n \vec{Z}_n \cdot
  \vec{\Sigma}_\mathrm{fiss}}{\lambda_d \vec{Z}_d \cdot \vec{\Sigma}_\mathrm{fiss}}
  e^{-(\lambda_n - \lambda_d) T_c},
\label{eq:ActRatio}
\end{equation}
%%%%%%%%%%%%%%%%%%%%%%%%%%%%%%%%
which is a direct measure of the total cooling time. One can correct
Eq.~\ref{eq:ActRatio} with higher order expansion terms to account for linear
fragments with large neutron-capture components, but this will create a dependence
on $T_\mathrm{irr}$. For large fast fluences, $\Phi$ must remain in
Eq.~\ref{eq:ActRatio} so as to account for fast fissions:
$\Phi_g \times (\vec{Z}^g_n\cdot\vec{\Sigma}^g_{\mathrm{fiss}})$, with an implied
sum over the neutron energy groups $g$.

As mentioned previously, the final value of $\vec{\Sigma}_\mathrm{fiss}$ is known
from a measurement of fissile isotopics. However, $\vec{\Sigma}_\mathrm{fiss}$
varies over the irradiation period. Therefore, one must average the weighted fission
cross-sections so as not to bias Eq.~\ref{eq:ActRatio} towards U or Pu fissions. The
averaging is conducted linearly over the fluence $\Phi$ because $T_\mathrm{irr}$ is
unknown. One can use the thermal fluence derived from Eq.~\ref{eq:FluenceD} as the
fluence endpoint and the initial value of $\vec{\Sigma}_\mathrm{fiss}$ reflected
natural uranium for our samples~\cite{Byerly:2015}. This fluence-averaged value
$\langle \vec{\Sigma}_\mathrm{fiss}\rangle_\Phi$ becomes a critical factor when
predicting fragments that have cumulative yields with large plutonium components.

Inverting Eq.~\ref{eq:ActRatio} reveals the cooling time diagnostic
%%%%%%%%%%%%%%%%%%%%%%%%%%%%%%%%
\begin{equation}
T_c = \frac{1}{\lambda_d-\lambda_n} \ln \bigg( \frac{\alpha_{n,d} \lambda_d
  \vec{Z}_d \cdot \langle\vec{\Sigma}_\mathrm{fiss}\rangle_\Phi}{\lambda_n
  \vec{Z}_n \cdot \langle\vec{\Sigma}_\mathrm{fiss}\rangle_\Phi} \bigg).
\label{eq:CoolingTime}
\end{equation}
%%%%%%%%%%%%%%%%%%%%%%%%%%%%%%%%
Due to the pole in Eq.~\ref{eq:CoolingTime}, two linear fragments with similar
decay constants $\lambda_n \simeq \lambda_d$, such as a ratio of $^{90}$Sr and
$^{137}$Cs, can produce large errors in the cooling time, but there are
theoretical methods for removing these~\cite{Cetnar2006640}. For fragments with
large cross-sections, one can expand Eq.~\ref{eq:LinearBateman} to
$\mathcal{O}( (\tilde{\lambda}T_\mathrm{irr})^2)$, but this introduces an
unverifiable value for $T_\mathrm{irr}$ and only corrects the cooling time by a
few percent.

%%%%%%%%%%%%%%%%%%%%%%%%%%%%%%%%%%%%%%%%%%%%%%%%%%%%%%%%%%%%%%%%%%%%%%%%%%%
\section{Verification \label{sec:Verification}}
In Sec.~\ref{sec:Fluence} and Sec.~\ref{sec:CoolingTime}, we listed diagnostics for
the thermal fluence and cooling time. These diagnostics were verified with the use
of the reactor simulation described in Sec.~\ref{sec:CoolingTime}. Over $70$ sample
cases were evaluated with layer $4$ nuclear data to determine the validity of the
analytical calculations. The cases spanned a range of reasonable values for the
thermal flux $\phi_t$, cooling time $T_c$, fast flux $\phi_f$, irradiation time
$T_\mathrm{irr}$, number of shutdowns $N_s$, and shutdown length $T_s$. The derived
values for $\Phi$ and $T_c$, using Eq.~\ref{eq:FluenceD} and Eq.~\ref{eq:CoolingTime},
were compared with those used as input to the simulation. We found that the only
parameter that affected the fluence diagnostic is the introduction of a fast flux
$\phi_f$ as it slightly increases the final $\rho$ and $\epsilon$ values, which could be
mistaken for a larger thermal fluence. Using the maximum expected fast flux, the
diagnostic of Eq.~\ref{eq:FluenceD} returned the input fluence within $\sim 0.5\%$ for
both the $^{235}$U/$^{238}$U and $^{236}$U/$^{235}$U ratios. The situation for the
cooling time diagnostic was much more complicated.

We used the following ratios for the cooling time diagnostic: $^{137}$Cs/$^{155}$Eu
($\alpha_1$), $^{137}$Cs/$^{125}$Sb ($\alpha_2$), and $^{155}$Eu/$^{125}$Sb
($\alpha_3$)~\footnote{Diagnostics using $^{85}$Kr were removed as it may have experienced volatile leakage.}.
The derived cooling time was found to vary with all major reactor parameters listed
above. As the total $\Phi_t$ increased, the errors on Eq.~\ref{eq:CoolingTime}
increased linearly, but this was shown to be mediated somewhat by the averaging of
$\vec{\Sigma}_\mathrm{fiss}$. The increase of $\phi_f$ created an underestimation of
$T_c$ proportional to the additional fast cumulative yields of the fragments used in
Eq.~\ref{eq:CoolingTime}. Increasing the cooling time served to decrease the errors
on all $T_c$ diagnostics as the deviation from end-of-cycle activity ratios becomes more
severe for longer $T_c$. Finally, the shutdown history is shown to have a very small
impact, in agreement with the derivation in Sec.~\ref{sec:CoolingTime}. The maximum
theoretical errors in percent for the expected reactor parameters and the largest
overall theoretical error are provided in Tab.~\ref{tab:ErrorAnalysis}.

%%%%%%%%%%%%%%%%%%%%%%%%%%
\begin{table}[h]
\centering
    \begin{tabular}{|c||c|c|c|c|c|}
        \hline 
	& \multicolumn{2}{c}{$\Phi$ Diagnostics} & \multicolumn{3}{|c|}{$T_c$ Diagnostics} \\ \hline
	& $\epsilon$ & $\rho$ & $\alpha_1$ & $\alpha_2$ & $\alpha_3$ \\ \hline \hline
	$\Phi_t$ & $\sim0\%$ & $\sim0\%$ & $3.86\%$ & $0.57\%$ & $-3.27\%$ \\ \hline
	$\phi_f$ & $0.54\%$ & $0.24\%$ & $-0.47\%$ & $-0.19\%$ & $-0.14\%$ \\ \hline
	$T_c$ & $0\%$ & $0\%$ & $-0.99\%$ & $-0.12\%$ & $0.89\%$ \\ \hline
	$N_s$ & $0\%$ & $0\%$ & $0.10\%$ & $0.01\%$ & $-0.10\%$ \\ \hline
	$T_s$ & $0\%$ & $0\%$ & $-0.17\%$ & $-0.16\%$ & $-0.14\%$ \\ \hline
	Overall & $\pm0.54\%$ & $\pm0.24\%$ & $\pm4.02\%$ & $\pm0.63\%$ & $\pm3.40\%$ \\ \hline
	    \end{tabular}
\caption{\label{tab:ErrorAnalysis} Theoretical errors for the fluence
  ($^{235}$U/$^{238}$U and $^{236}$U/$^{235}$U ratios) and cooling
  time ($\alpha_1=\,^{137}$Cs/$^{155}$Eu, $\alpha_2=\,^{137}$Cs/$^{125}$Sb,
  and $\alpha_3=\,^{155}$Eu/$^{125}$Sb ratios) diagnostics given by
  Eq.~\ref{eq:FluenceD} and Eq.~\ref{eq:CoolingTime}. Each cell shows
  the maximum expected error over a particular reactor parameter (the
  thermal fluence $\Phi_t$, fast flux $\phi_f$, cooling time $T_c$,
  number of shutdowns $N_s$, and length of shutdowns $T_s$) range.
  The overall theoretical error is the individual errors summed in
  quadrature, which provides a conservative maximum.}
\end{table}
%%%%%%%%%%%%%%%%%%%%%%%%%%

Overall, from Tab.~\ref{tab:ErrorAnalysis}, one can see that the diagnostics derived
in Eq.~\ref{eq:FluenceD} for the fluence have extremely small theoretical errors and
one can expect the correct fluence within $\sim0.5\%$. For the cooling time
diagnostic, the theoretical errors are more substantial as the fragment systems are
more complex. Overall, one can expect the correct cooling time within $\sim4\%$,
$\sim0.6\%$, and $\sim3.4\%$ for the $^{137}$Cs/$^{155}$Eu, $^{137}$Cs/$^{125}$Sb,
and $^{155}$Eu/$^{125}$Sb ratios, respectively. The linear-averaging in
Sec.~\ref{sec:CoolingTime} returned the lowest errors, but ignores the quadratic
behavior of $^{239}$Pu at low burnup. We verified that removing $^{239}$Pu fissions
and calculating Eq.~\ref{eq:CoolingTime} to
$\mathcal{O}((\tilde{\lambda}T_\mathrm{irr})^2)$ can effectively eliminate these
errors. We note that these errors are strictly theoretical and contain no systematic errors, such
as fractionation or experimental uncertainties. We have also calculated the expected
$^{239}$Pu abundance using a similar analytical method with errors of $\sim0.25\%$, but
this requires knowledge of many reactor parameters, so we have excluded it from our
analysis. The theory errors of Tab.~\ref{tab:ErrorAnalysis} are lower than the
experimental measurement errors. With these notes in mind, we use these diagnostics
to determine the thermal fluence and extract information about systematic errors
from three cooling time diagnostics.

%%%%%%%%%%%%%%%%%%%%%%%%%%%%%%%%%%%%%%%%%%%%%%%%%%%%%%%%%%%%%%%%%%%%%%%%%%%
\section{Experimental Application \label{sec:Application}}
Ten archived samples were analyzed for their U and Pu isotopics, as well as the
activities of several fission fragments. The actinides were separated and measured
as described in Ref.~\cite{Byerly:2015}. In short, U metal or UO$_3$ samples are
dissolved in HNO$_3$, then loaded and separated on anion exchange columns to achieve
separation of Pu from U. Isotope ratios and isotope dilution measurements were
determined by TIMS. Fission fragments were measured by gamma
spectrometry~\cite{Tandon:2009}. Samples H and K were in UO$_3$ form, while the
remainder were uranium metal.

\begin{figure}[h]
\centering
\includegraphics[width=\columnwidth]{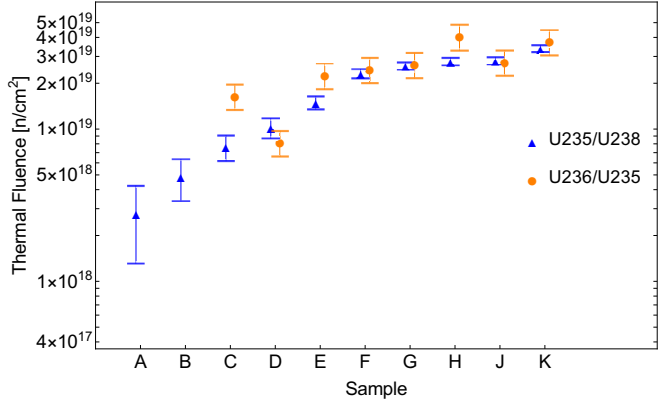}
\caption{\label{fig:PhiDetermination} The derived thermal fluence for ten
  experimental samples. The derivation for the thermal fluence used both the
  $^{235/238}$U (triangles) and $^{236/235}$U (circles) ratios and these methods
  are self-consistent in higher fluence samples. The $^{236/235}$U diagnostic
  could not be used in samples with trace $^{236}$U amounts. Errors are the
  $1\sigma$ errors from experimental measurements and theoretical estimates
  summed in quadrature. Color online.}
\end{figure}

Both fluence diagnostic methods were attempted, but discrepancies were observed
between the $^{236/235}$U and $^{235/238}$U ratios in very-low burnup cases as shown
in Fig.~\ref{fig:PhiDetermination}.
%%%%%%%%%%%%%%%%%%%%%%
%%%%%%%%%%%%%%%%%%%%%%
The fluences determined in samples D through K were all nearly self-consistent.
Sample C reported fluences that deviate more strongly. Samples A and B were
contaminated with $^{236}$U memory effects, so their values were not used. The
chemical analyses of the remaining samples were conducted at a later time,
correcting the $^{236}$U issue. Overall, it appears that our method of extracting
the thermal fluence via Eq.~\ref{eq:FluenceD} is accurate and self-consistent
for the majority of samples with $\Phi\geq10^{19}\,\mathrm{n/cm^2}$. Below this
limit, the low concentrations of $^{236}$U created experimental difficulties in
acquiring the fluence with multiple methods. Thus, one can determine the thermal
fluence with two independent diagnostics in samples with appreciable amounts of
$^{236}$U, but must rely solely on the $^{235/238}$U ratio in extremely low-burnup
samples with trace levels of $^{236}$U. The $\epsilon$ diagnostic is only
valid when $\epsilon_0$ is known, so the $\rho$ diagnostic should be used if
enough $^{236}$U is present. The average error between the two diagnostics was
$19.9\%$.

\begin{figure}[h]
\centering
\includegraphics[width=\columnwidth]{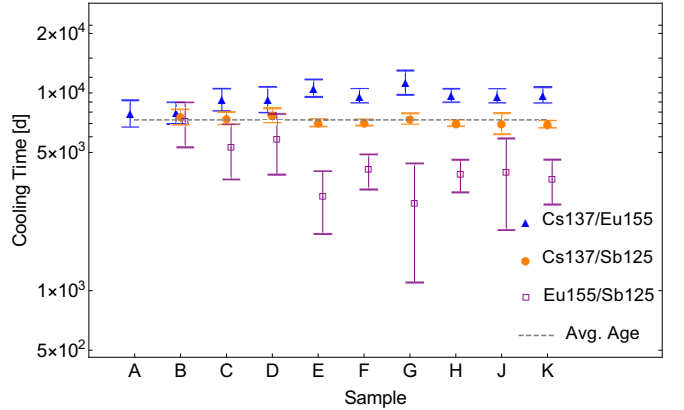}
\caption{\label{fig:CoolingTime} The derived cooling times for ten
  experimental samples. The derivation of the cooling time used $3$
  ratios, $\alpha_1$ (triangles), $\alpha_2$ (circles), and $\alpha_3$
  (open squares). See text for definition. The average sample age of
  $20\,\mathrm{yr}$ is shown for comparison with the derived values.
  The errors are the $1\sigma$ errors from the experimental
  measurements and theoretical estimates summed in quadrature. The
  disagreement between the diagnostics indicates the presence of
  systematic errors, which could be attributed to fractionation.
  Color online.}
\end{figure}

In determining the total cooling time, we used the ratios identified in
Sec.~\ref{sec:Verification}. Figure~\ref{fig:CoolingTime} illustrates the
agreement and tension between the different diagnostics. A few samples performed
relatively well, but most demonstrated disagreement between the three cooling
time diagnostics.
%%%%%%%%%%%%%%%%%%%%%%
%%%%%%%%%%%%%%%%%%%%%%
In particular, the $^{155}$Eu-based determinations of $T_c$ showed disagreement with
the $^{137}$Cs/$^{125}$Sb ratio as the inferred fluence rises. Leakage of volatile
fission fragments, such as $^{85}$Kr, can occur at the $\gtrsim 13\%$ level in PWR
fuels~\cite{Metz:2013} so these ratios were removed. A portion of the bias from
$^{155}$Eu-based diagnostics can be explained by the over-estimation of the
$^{239}$Pu-component when linearly averaging $\vec{\Sigma}_\mathrm{fiss}$ and the
need to compute $T_c$ to second order, but these errors will only approach those in
Tab.~\ref{tab:ErrorAnalysis}. The $^{137}$Cs/$^{125}$Sb diagnostic seemed to match
the average reported sample age of $20\,\mathrm{yr}$. The $^{125}$Sb abundance was
not measured in sample A. The average diagnostic discrepancy was found to be
$\sim37\%$ between the $^{155}$Eu-based diagnostics and the $^{137}$Cs/$^{125}$Sb
ratio. The use of multiple $T_c$ diagnostics allows one to detect the presence of
systematic errors, such as fractionation, when diagnostics do not agree and a single
consistent cooling time when they do. This technique must be used in the very low
burnup regime as traditional same-species ratios are impractical.

%%%%%%%%%%%%%%%%%%%%%%%%%%%%%%%%%%%%%%%%%%%%%%%%%%%%%%%%%%%%%%%%%%%%%%%%%%%
\section{Conclusion \label{sec:Conclusion}}
The work conducted here demonstrates that the thermal fluence can be determined in
low burnup samples using the $^{235}$U/$^{238}$U and $^{236}$U/$^{235}$U ratios.
These ratios are common fluence diagnostics, which were verified with detailed
reactor simulations and then experimentally demonstrated to be accurate and self-
consistent when enough $^{236}$U is produced above the detection threshold. The
average discrepancy between the two fluence diagnostics in our low burnup samples
was $19.9\%$ for $\Phi>10^{19}\,\mathrm{n/cm^2/sec}$.

The low burnup of our reactor samples required new cooling time diagnostics to be
derived, as the concentrations of standard diagnostic tags are below detection
thresholds. The new cooling time diagnostics utilized simple linear fission
fragments with long half-lives and considerable fission yields. Four such fragments
were identified and the derived diagnostics were verified in low burnup scenarios.
The experimentally determined cooling times were shown to be consistent in some
samples, but varied by $\sim37\%$ on average. In addition, leakage of volatile gases
invalidated the diagnostics using $^{85}$Kr. Overall, the $^{137}$Cs/$^{125}$Sb
ratio seemed to agree with the average sample age across all samples. Differing
results for the cooling time, as measured by several diagnostics, proved to be
indicative of systematic errors, such as fractionation, even in the very low burnup
regime.

The fluence and cooling time derivation should be conducted in tandem, where the
$\Phi$ determination would be used to derive
$\langle\vec{\Sigma}_\mathrm{fiss}\rangle_\Phi$ and verify that the sample has a
burnup low enough to validate the simple analytical expressions for $T_c$. This
work provides verification of fluence diagnostics and new cooling time diagnostic
techniques to determine the presence of systematic errors in low burnup samples,
both of which have applications in non-proliferation and verification.

%%%%%%%%%%%%%%%%%%%%%%%%%%%%%%%%%%%%%%%%%%%%%%%%%%%%%%%%%%%%%%%%%%%

\acknowledgements
We would like to thank the analytical chemistry team: L. Colletti, E. Lujan, K. Garduno, T. Hahn, L. Walker, A. Lesiak, P. Martinez, F. Stanley, R. Keller, M. Thomas, K. Spencer, L. Townsend, D. Klundt, D. Decker, and D. Martinez. Los Alamos National Laboratory supported this work through LDRD funding. This publication is LA-UR-16-26969.

%\pagebreak
%%%%%%%%%%%%%%%%%%%%%%%%%%%%%%%%%%%%%%%%%%%%%%%%%%%%%%%%%%%%%%%%%%%
\bibliographystyle{apsrev} \bibliography{LinearDiagnostics.bib}
%%%%%%%%%%%%%%%%%%%%%%%%%%%%%%%%%%%%%%%%%%%%%%%%%%%%%%%%%%%%%%%%%%%

\end{document}